\pdfoutput=1
\documentclass[11pt]{article}

\usepackage[preprint]{acl}

\usepackage[T1]{fontenc}
\usepackage[utf8]{inputenc}

\usepackage{microtype}

\usepackage{inconsolata}

\usepackage{graphicx}
\usepackage{algorithm}
\usepackage{algpseudocode}
\usepackage{amssymb}
\usepackage{amsmath}
\usepackage[most]{tcolorbox}
\usepackage{pgfplots}
\usepackage{enumitem}
\usepackage{subfig}
\usepackage{colortbl}  
\usepackage{xcolor}    
\usepackage{times}
\usepackage{arydshln}
\usepackage{latexsym}
\usepackage{caption}
\usepackage{pifont}
\usepackage{booktabs}
\title{Ramp Up NTT in Record Time using GPU-Accelerated Algorithms and LLM-based Code Generation}

\author{
\textbf{Yu Cui\textsuperscript{1,2}\thanks{Co-first authors contributed equally.}},
\textbf{Hang Fu\textsuperscript{1}\footnotemark[1]},
\textbf{Licheng Wang\textsuperscript{1}\thanks{Corresponding authors.}},
\textbf{Haibin Zhang\textsuperscript{2}\footnotemark[2]}
\\
\textsuperscript{1}Beijing Institute of Technology \\
\textsuperscript{2}Yangtze Delta Region Institute of Tsinghua University, Zhejiang \\
\texttt{\{cuiyu, fuhang, lcwang\}@bit.edu.cn, bchainzhang@aliyun.com}
}

\begin{document}
\maketitle
\begin{abstract}
 Homomorphic encryption (HE) is a core building block in privacy-preserving machine learning (PPML), but HE is also widely known as its efficiency bottleneck. Therefore, many GPU-accelerated cryptographic schemes have been proposed to improve the performance of HE. However, these methods often require complex modifications tailored to specific algorithms and are tightly coupled with specific GPU and operating systems. It is interesting to ask how to generally offer more practical GPU-accelerated cryptographic algorithm implementations. Given the powerful code generation capabilities of large language models (LLMs), we aim to explore their potential to automatically generate practical GPU-friendly algorithm code using CPU-friendly code. In this paper, we focus on number theoretic transform (NTT)---the core mechanism of HE. We first develop and optimize a GPU-friendly NTT (GNTT) family that exploits PyTorch's fast matrix computation and precomputation, achieving an approximately $62 \times$ speedup---a significant boost over existing ones. Then we explore GPU-friendly code generation using various LLMs, including DeepSeek-R1, OpenAI o1 and o3-mini. We discover many interesting findings throughout the process. For instance, somewhat surprisingly, our experiments demonstrate that DeepSeek-R1 significantly outperforms OpenAI o3-mini and o1, but still cannot beat our optimized protocol. The findings provide valuable insights for turbocharging PPML and enhancing code generation capabilities of LLMs. Codes are available at: \url{https://github.com/LMPC-Lab/GenGPUCrypto}.

\end{abstract}

\begin{table*}[t]
\centering
\renewcommand{\arraystretch}{1.2}
\setlength{\tabcolsep}{15pt}
\scalebox{0.71}{
\begin{tabular}{cll}
\toprule
\textbf{Algorithm} & \textbf{Implementation} & \textbf{Precomputation}\\
\midrule
\midrule
GNTT1& Matrix parallel computation using PyTorch & Serial computing\\
GNTT2 & LUT-based acceleration & Parallel and serial computing\\
GNTT3 & Using CuPy to reduce memory consumption & Parallel and serial computing\\
GNTT4 & Utilizing CuPy to support large parameters for HE& Parallel and serial computing\\
\bottomrule
\end{tabular}
}
\caption{GNTT family algorithms. The four algorithms are similar in their core NTT operations, with differences lying in the precomputation and optimizations in implementation details. GNTT2 has the optimal execution efficiency, while GNTT3 and GNTT4 are enhanced based on GNTT2.}
\label{tab: GNTT-family}
\end{table*}

\section{Introduction}
In recent years, aimed at ensuring the privacy of machine learning, significant research efforts have focused on achieving privacy-preserving machine learning (PPML) \citep{Ng2023SoK}. The core of these PPML solutions lies the utilization of homomorphic encryption (HE) and secure multiparty computation (MPC) \citep{Pang2024BOLT, Choi2024Touch, rho2024encryption}. While ensuring data confidentiality, the use of these cryptographic algorithms introduces substantial computational overhead. To address this efficiency challenge, researchers have proposed numerous optimization approaches to accelerate model training and inference over ciphertexts \citep{Frimpong2024GuardML, Huang2024GeniBatch, Yuan2024MDML}. Similar to PPML, realizing privacy protection for large language models (LLMs) has become increasingly significant \citep{Das2025survey}. Moreover, \citet{zhao2024privacy} highlights the necessity of constructing privacy-preserving LLMs, which can maintain performance.

For the efficiency optimization of both PPML and privacy-preserving LLMs, achieving GPU acceleration for HE and MPC is the critical task. However, existing GPU-accelerated cryptographic schemes \citep{shen2024BFV, moon2024thor} require complex modifications and are tightly coupled with the internal structure of GPUs, operating system and even hardware, lacking generality and portability. For fully HE with hardware architecture, \textbf{limited generality is an obstacle for further accelerating polynomial computation}, since it is difficult for specific-designed hardware accelerators to maintain optimal acceleration as the degree of the polynomials changes \citep{zhang2024sok}. 

Therefore, a better strategy is to bypass the underlying architecture and construct GPU-accelerated cryptographic algorithms from the perspective of high-level language frameworks, thereby breaking through the limitations of generality. \citet{Tan2021CryptGPU} exploits PyTorch \citep{Paszke2019pytorch} to construct a MPC framework for PPML, which implements the entire cryptographic operations on GPU. They argue that directly translating CPU-based code to the GPU is difficult to improve performance immediately due to the distinct designs in architecture between the CPU and GPU. Their research also emphasizes the importance to build GPU-friendly cryptography. Many researches aimed at realizing GPU-friendly algorithms for PPML has concentrated on MPC, and there remains a notable gap in studies exploring the parallelization for HE. Therefore, it is interesting to ask how to generally offer more practical GPU-accelerated HE implementations.

Motivated by the powerful ability of LLMs in code generation recently \citep{yan2024codescope}, we introduce an innovative research problem: \textbf{Do LLMs have excellent GPU-friendly cryptographic code generation capability?}
In this paper, we take number theoretic transform (NTT) \citep{Agarwal1975NTT} as the core research object, since underlying polynomial computation with high computational complexity is one of the two root factors of fully HE's inefficiency, and NTT is crucial for optimizing these polynomial operations \citep{zhang2024sok}. Therefore, NTT's efficiency directly influences the performance of PPML. We require LLMs to generate GPU-friendly NTT code built on PyTorch, according to the given CPU-friendly code of Fast-NTT \citep{Satriawan2024BeginnerGuide}, and benchmark the generation capability of LLMs.

We optimize and implement a practical GPU-friendly NTT (GNTT) family that leverages PyTorch’s efficient matrix computation and precomputation to significantly boost the computational speed. The GNTT family consists of four distinct algorithm instances (GNTT1–GNTT4). Additionally, we exploit CuPy \citep{nishino2017cupy} to alleviate issues related to floating-point type conversions and high memory consumption. A comparative analysis of the implementation details for each GNTT instance is provided in \autoref{tab: GNTT-family}.

For models to study, we choose DeepSeek-R1 \citep{guo2025deepseek}, OpenAI o3-mini \footnote{https://openai.com/index/openai-o3-mini} and o1 \citep{jaech2024openai}, which demonstrate outstanding performance in mathematics and programming. We conduct a thorough analysis of the generated codes from an algorithmic theory perspective. Furthermore, the experimental results have shown that GNTT realizes nearly a around $62 \times$ speedup under ideal conditions, compared to the CPU-friendly Fast-NTT. Overall, \textbf{DeepSeek-R1 significantly outperforms OpenAI o3-mini and o1, but it still far from reaching the level of GNTT}. To sum up, the contributions of this paper are as follows: 
\begin{itemize}[leftmargin=*, noitemsep]

\item \textbf{Innovative research problem}: To our knowledgae, this paper is the first work to introduce the research problem of GPU-friendly cryptographic code generation capabilities in LLMs, which is essential to generally realize more practical GPU-accelerated HE.

\item \textbf{GNTT family}: We first develop and optimize GNTT family leveraging PyTorch's fast matrix computation and precomputation. To improve the performance of GNTT, lookup table (LUT) and CuPy serve as innovative solutions to accelerate the precomputation and reduce memory consumption, respectively. By utilizing GPU-friendly calculating operation, GNTT is far superior to Fast-NTT. 

\item \textbf{Comprehensive theory analysis}:
We conduct a comprehensive analysis of the algorithms generated by LLMs from the theoretical perspective of cryptography, providing a detailed comparison of the differences between each algorithm.

\item \textbf{Experimental findings}: The experimental results indicate that existing LLMs show significant variations and exhibit defects in their GPU acceleration solutions. The findings offer significant insights that can drive future advancements in PPML and contribute to the enhancement of code generation capabilities in LLMs.

\end{itemize}

\begin{figure*}[t!]
    \centering
    \includegraphics[width=1.0\linewidth]{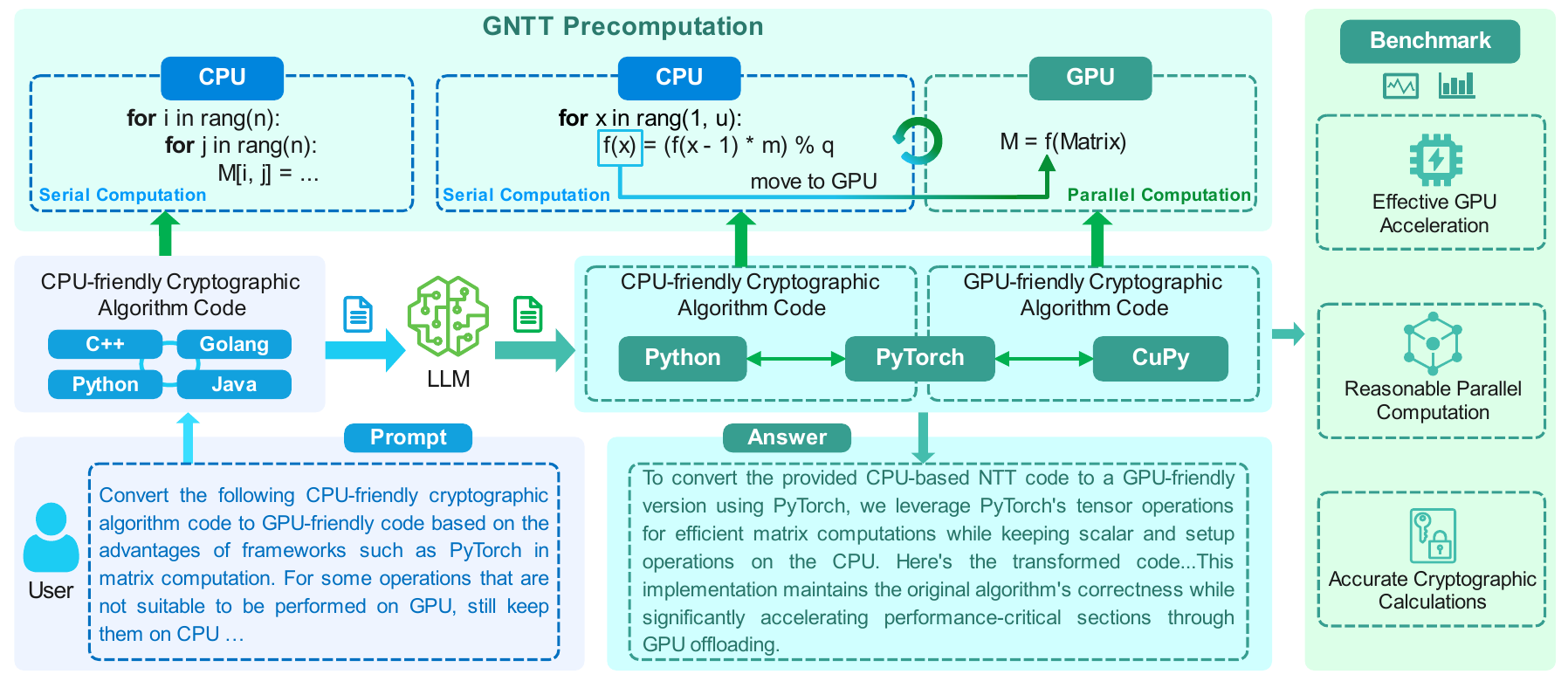}
    \caption{The framework of GPU-Accelerated cryptographic algorithm code generation based on LLMs and its benchmarking. We provide an example of algorithmic code generation on how to convert the core CPU-friendly code for GNTT precomputation into a code that takes into account the computational advantages of both CPU and GPU. This example actually illustrates the core difference between GNTT1 and GNTT2.}
    \label{fig:ecosystem}   
\end{figure*}

\section{Approach}
\subsection{Framework Overview}
The core framework for GPU-accelerated cryptographic algorithm code generation leveraging LLMs is illustrated in \autoref{fig:ecosystem}. We refer to \citep{wu2024need} to formalize this framework as
\begin{equation}
    (g, t) = F(c, d, p(x) | \theta),
\end{equation}
where $d$ represents various types of CPU and GPU, and $F$ is the LLM parameterized by $\theta$. The code generation prompt $p$ includes the parameters required for executing cryptographic algorithms (e.g., $q$ for the NTT algorithm in Section \ref{sec: algorithm design}). Given a CPU-friendly code $c$ written in any programming language, the LLM generates the corresponding GPU-friendly code via code translation. The LLM then utilizes external tools \citep{Schick2023Toolformer} to execute code $g$ in an environment equipped with $d$, benchmarking the algorithm's execution time $t$. $g$ primarily utilizes frameworks like PyTorch and CuPy, which enable high-speed matrix parallel computation. This formalized expression intuitively demonstrates that the performance of cryptographic algorithms is closely related to $d$, and it exhibits a high degree of generality for $d$ with various configurations.

\subsection{Code Generation Benchmark}
To evaluate the quality of GPU-accelerated code generated by LLMs, we propose a benchmark as follows:

\begin{itemize}[leftmargin=*, noitemsep]
\item \textbf{Effective GPU acceleration}: The GPU-accelerated code should exhibit a substantial increase in execution speed compared to the original CPU-friendly code. Moreover, the conversion process should go beyond basic transformations, such as directly converting scalars to tensors, and should also encompass the conversion of cryptographic algorithms themselves. This will ensure the potential of matrix parallel computing is fully leveraged.

\item \textbf{Reasonable parallel computation}: 
Admittedly, not all computations are suited for GPU acceleration. Some CPU-friendly computations that are difficult to parallelize, should maintain their original computational logic. If all computational operations are forcibly converted using PyTorch, it may actually slow down the execution speed. Therefore, the code generated by LLMs should embody both CPU-friendly and GPU-friendly code, balancing the advantages of both to achieve maximum speed optimization.

\item \textbf{Accurate cryptographic calculations}:
The correctness of the execution result of the converted GPU-accelerated cryptographic algorithm code should not be affected.

\end{itemize}

\section{Comprehensive Analysis}
\label{sec: algorithm design}
This section presents a comprehensive analysis of the GNTT family and the code generation capabilities of LLMs. The notation and terminology for NTT-based negative-wrapped convolution are detailed in \autoref{sec:notation}.

\definecolor{mgreen}{rgb}{0, 0.6, 0.3}
\subsection{GNTT Family}
The GNTT family comprises four distinct GPU-friendly NTT algorithm instances, optimizing computational speed by leveraging PyTorch for efficient matrix computations. In the algorithm tables later, for ease of distinction, operations executed on the GPU will be highlighted in \textcolor{mgreen}{green}.

\definecolor{newred}{RGB}{254,115,75}

\begin{table*}[t]
\centering
\renewcommand{\arraystretch}{1.1}
\scalebox{0.95}{
\begin{tabular}{lcccc}
\toprule
\textbf{Algorithm} & \textbf{FNTT-o1} & \textbf{FNTT-o3} & \textbf{FNTT-DS} & \textbf{GNTT (Baseline)} \\
\midrule
\midrule
High GPU friendliness & \textcolor{newred}{\ding{55}}  & \textcolor{newred}{\ding{55}} & \textcolor{newred}{\ding{55}} & \textcolor{mgreen}{\ding{51}} \\
Thoughts on precomputation & \textcolor{newred}{\ding{55}} & \textcolor{mgreen}{\ding{51}} & \textcolor{mgreen}{\ding{51}} & \textcolor{mgreen}{\ding{51}} \\
Correct precomputation & \textcolor{newred}{\ding{55}} & \textcolor{mgreen}{\ding{51}} & \textcolor{newred}{\ding{55}} & \textcolor{mgreen}{\ding{51}} \\
Reasonable parallel computation & \textcolor{newred}{\ding{55}}  & \textcolor{newred}{\ding{55}} & \textcolor{mgreen}{\ding{51}} & \textcolor{mgreen}{\ding{51}} \\
Application of mathematical skills & \textcolor{newred}{\ding{55}}  & \textcolor{newred}{\ding{55}} & \textcolor{mgreen}{\ding{51}} & \textcolor{newred}{\ding{55}} \\
\bottomrule
\end{tabular}
}
\caption{The comprehensive comparative analysis of GPU-friendly NTT algorithm codes from the theoretical perspective
of cryptographic algorithms.}
\label{tab: Comparison}
\end{table*}

\subsubsection{GNTT1}
Following the formulations provided in \citep{Satriawan2024BeginnerGuide}, equations (13)$\sim$(18), the NTT can be rewritten in matrix form as:
\begin{equation}
\boldsymbol{a'} = \mathbf{W}_{\text{NTT}} \cdot \boldsymbol{a} \pmod{q},
\end{equation}
where the transformation matrix $\mathbf{W}_{\text{NTT}}$ is defined as:
\begin{equation}
\resizebox{1\linewidth}{!}{$
\mathbf{W}_{\text{NTT}} = \left(\begin{array}{cccc}
\psi^0 & \psi^1 & \cdots & \psi^{n-1} \\
\psi^0 & \psi^3 & \cdots & \psi^{3 \cdot (n-1)} \\
\vdots & \vdots & \ddots & \vdots \\
\psi^0 & \psi^{(2n-1)} & \cdots  & \psi^{(2n-1)\cdot (n-1)}
\end{array}\right).
$}
\end{equation}
Similarly, the inverse transformation can be expressed as:
\begin{equation}
\boldsymbol{c} = n^{-1} \cdot \mathbf{W}_{\text{INTT}} \cdot \boldsymbol{c'} \pmod{q},
\end{equation}
where the inverse transformation matrix $\mathbf{W}_{\text{INTT}}$ is given by:
\begin{equation}
\resizebox{1\linewidth}{!}{$
\mathbf{W}_{\text{INTT}} = \left(\begin{array}{cccc}
\psi^{-0} & \psi^{-0} & \cdots & \psi^{-0} \\
\psi^{-1} & \psi^{-3} & \cdots & \psi^{-(2n-1)} \\
\vdots & \vdots & \ddots & \vdots \\
\psi^{-(n-1)} & \psi^{-3 \cdot (n-1)} & \cdots  & \psi^{-(2n-1)\cdot (n-1)}
\end{array}\right).
$}
\end{equation}

This matrix formulation provides a structured representation of NTT and its inverse, making it well-suited for implementation in PyTorch. To demonstrate this approach, Algorithm \ref{alg:GNTT111} provides a full implementation of the method. While the algorithm is computationally inefficient in terms of raw performance, it effectively illustrates the NTT algorithmic flow and serves as a valuable reference for understanding the key steps in NTT computations.

GNTT1 leverages the efficiency of tensor computations on GPU to accelerate NTT. Key operations, such as matrix multiplication and element-wise multiplication, are efficiently executed on the GPU. However, the calculation of matrix entries \( \psi^{2ij + j} \), which requires \( O(n^2) \) exponentiations, is not suitable for GPU parallelization and is instead performed on the CPU. This step introduces significant computational overhead and reduces the algorithm's overall efficiency. Further optimization will be addressed in subsequent algorithms.

\begin{algorithm}[H]
\caption{GNTT1}
\label{alg:GNTT111}
\raggedright
\textbf{Input:} Polynomials $\boldsymbol{a}, \boldsymbol{b}$ of length $n$, prime modulus $q$, a primitive $2n$-th root of unity $\psi$, d. \\
\textbf{Output:} $\boldsymbol{c} = INTT(NTT(\boldsymbol{a}) \odot NTT(\boldsymbol{b}))$.

\begin{algorithmic}[1]
    \State {Initialize $\mathbf{W}_{\text{NTT}}$ and $\mathbf{W}_{\text{INTT}}$, size $n \times n$.}

    \For{$i,j$} 
        \State $\mathbf{W}_{\text{NTT}}[i,j] \gets \psi^{2ij + j} \mod q$
    \EndFor
    \State Move matrix to device: $\mathbf{W}_{\text{NTT}} \gets \mathbf{W}_{\text{NTT}}.to(\text{cuda})$

    \State \textcolor{mgreen}{Compute NTT: $\boldsymbol{a'} \gets \mathbf{W}_{\text{NTT}} \cdot \boldsymbol{a} \mod q$}
    \State \textcolor{mgreen}{Compute NTT: $\boldsymbol{b'} \gets \mathbf{W}_{\text{NTT}} \cdot \boldsymbol{b} \mod q$}

    \State \textcolor{mgreen}{$\boldsymbol{c'} \gets (\boldsymbol{a'} \odot \boldsymbol{b'}) \mod q$}

    \State Compute inverse: $\psi^{-1}$ and $n^{-1}$
    \For{$i,j$} 
        \State $\mathbf{W}_{\text{INTT}}[i,j] \gets (\psi^{-1})^{2ij + i} \mod q$
    \EndFor
    \State Move matrix to device: $\mathbf{W}_{\text{INTT}} \gets \mathbf{W}_{\text{INTT}}.to(\text{cuda})$
    
    \State \textcolor{mgreen}{Compute INTT:} \parbox[t]{\linewidth}{\textcolor{mgreen}{$\boldsymbol{c} \gets n^{-1} \cdot \mathbf{W}_{\text{INTT}} \cdot \boldsymbol{c'} \mod q$}}
    
    \State \textbf{Return:} $\boldsymbol{c}$

\end{algorithmic}
\end{algorithm}

\subsubsection{GNTT2}
To address the computational complexity of calculating powers of the primitive $2n$-th root of unity, we note that the computation depends solely on $n$ and $q$, making it suitable for precomputation. This allows us to transfer the computation to the offline phase. Specifically, we enhance the matrix construction process from GNTT1 by precomputing and storing the power values in LUTs. This optimization enables the online phase to concentrate on low-overhead tensor operations, such as matrix multiplication, which are computationally efficient.

In addition to LUT precomputation, GNTT2 further optimizes matrix construction by utilizing PyTorch’s broadcasting mechanism. Instead of computing each matrix entry individually, the transformation matrices $\mathbf{W}_{\text{NTT}}$ and $\mathbf{W}_{\text{INTT}}$ are directly constructed by indexing the LUTs, as illustrated in the green-highlighted sections of Algorithm~\ref{alg:GNTT222}. This design capitalizes on the efficiency of PyTorch’s tensor operations, where broadcasting maps precomputed values across the matrix in a single operation, thereby reducing memory access overhead.

\begin{algorithm}[H]
\caption{GNTT2}
\label{alg:GNTT222}
\raggedright
\textbf{Input:} $\boldsymbol{a}, \boldsymbol{b}$, $n$, $q$, $\psi$, $\psi^{-1}$, $d$. \\
\textbf{Output:} $\boldsymbol{c}$.

\begin{algorithmic}[1]
    \State \textbf{Offline Computation: Matrix Construction}
    \State $i,j \gets \text{range}(0, n)$
    \State $\text{exponents}[i, j] \gets 2ij + j$
    \State Precompute the LUT for powers of $\psi \bmod q$:  
    \For{$k = 0$ to $\max(\text{exponents})$}
        \State $\text{LUT}_{\psi}[k] \gets \psi^k \bmod q$
    \EndFor
    \State \resizebox{0.44\textwidth}{!}{Precompute the LUT for powers of $\psi^{-1} \bmod q$:}  
    \For{$k = 0$ to $\max(\text{exponents})$}
        \State $\text{LUT}_{\psi^{-1}}[k] \gets (\psi^{-1})^k \bmod q$
    \EndFor
    \State \parbox[t]{\linewidth}{
        \textcolor{mgreen}{$\mathbf{W}_{\text{NTT}} \gets \text{LUT}_{\psi}[\text{exponents}]$}\\[1mm]
        \textcolor{mgreen}{$\mathbf{W}_{\text{INTT}} \gets \text{LUT}_{\psi^{-1}}[\text{exponents}]$}
    } 
    
    \State \textbf{Online Computation: NTT \& INTT}
    \State \Comment{The online phase follows the methodology presented in GNTT1.}
    \State \textbf{Return:} $\boldsymbol{c}$
\end{algorithmic}
\end{algorithm}

\subsubsection{GNTT3 and GNTT4}
The core optimization of GNTT3 and GNTT4 lies in the use of CuPy to address the issues encountered with PyTorch during computation. Building upon GNTT2, GNTT3 utilizes CuPy to reduce memory consumption and avoids the cumbersome conversion between floating-point and integer types during matrix-vector multiplication. Additionally, when the parameters $n$ and $q$ are very large, GNTT2 may experience overflow in the modular operation of INTT. To effectively alleviate this problem, we transfer the result tensor back to the CPU and perform the modular operation using CuPy's modular function.

\subsection{GPU-friendly NTT generated by LLMs}
This section evaluates the ability of LLMs to convert CPU-friendly Fast-NTT into efficient, PyTorch-based GPU-friendly versions. We provide each model with the same source code and prompt them to generate GPU-accelerated implementations. We analyze the responses (FNTT-o1, FNTT-o3, FNTT-DS) from three models: OpenAI o1, o3-mini and DeepSeek-R1, all of which successfully produce corresponding PyTorch implementations accompanied by theoretical explanations. 
The comparison of these algorithm codes is shown in \autoref{tab: Comparison}. A detailed analysis of their outputs allows us to evaluate whether they meet the properties outlined in our proposed benchmark, focusing on GPU acceleration, parallelism, and correctness.

\subsubsection{FNTT-o1}
FNTT-CPU is implemented using pure Python loops and sequential computations, making it well-suited for small-scale operations characterized by strong data dependencies, such as modular exponentiation and modular inversion. Because these operations involve limited data and exhibit inherent sequential dependencies, executing them on the CPU circumvents the overhead associated with data transfers and kernel launches, thereby preserving high efficiency.

In contrast, FNTT-o1 converts all operations into tensor computations and offloads them to the GPU. However, it retains the sequential algorithmic structure inherited from FNTT-CPU, which was originally designed for CPU execution. Consequently, although operations such as modular exponentiation and butterfly computations are performed using tensors, they are still processed in a step-by-step manner. This inherent sequential execution prevents the algorithm from fully leveraging GPU parallelism. While FNTT-o1 benefits from GPU acceleration for massively parallel tasks, indiscriminately offloading all operations, including those more efficiently executed on the CPU, introduces unnecessary computational overhead and ultimately degrades overall performance.

\subsubsection{FNTT-DS}
Unlike FNTT-o1, FNTT-DS restructures the algorithm to fully exploit GPU parallelism by representing computations as batched tensor operations rather than relying on the sequential control flow inherited from CPU-friendly designs. In FNTT-DS, modular exponentiation is reformulated to update entire tensors in parallel. Specifically, a boolean mask identifies positions in the exponent tensor where the exponent is odd, enabling simultaneous updates to the corresponding elements of the result tensor with the current base value. Although the implementation still uses a while loop to iterate until all exponent values are reduced to zero, each iteration processes the entire tensor at once through a single kernel launch. This vectorized approach eliminates the need for launching separate kernels for individual element updates, reducing kernel launch overhead. Additionally, FNTT-DS subtly applies Fermat’s Little Theorem to compute modular inversion efficiently. Rather than using an iterative procedure, it employs the identity $x^{-1} \equiv x^{q-2} \mod q$ to perform modular inversion by invoking the modular exponentiation routine once.

The NTT algorithm in FNTT-DS uses a for loop to manage the transformation stages, but the actual computation within each stage, including rotation factor calculation and butterfly operations, is performed in parallel through batched tensor operations. This approach fully utilizes the GPU's parallelism, processing the entire dataset concurrently in each stage, leading to significant performance improvements. The design reflects the principles of reasonable parallel computing, combining sequential control with extensive parallel execution to accelerate computationally demanding tasks.

It is worth noting that DeepSeek-R1 intended to precompute the indices for order-reversal sorting and the rotation factors in the NTT process to optimize efficiency. The model attempted to eliminate redundant calculations and improve performance by precomputing these values. Unfortunately, order-reversal is tightly coupled with the specific polynomial being processed, making it ineffective as a true precomputation step. Furthermore, by calculating the rotation factors during each iteration, this precomputation strategy ultimately fails. Instead of computing the factors once and reusing them, they are recalculated in each loop iteration, which limits the ability to fully exploit parallelization and optimization.

\subsubsection{FNTT-o3}
The key improvement in FNTT-o3 is the precomputation of rotation factors. Unlike other models, which recalculate rotation factors in each iteration, FNTT-o3 computes them once before the main loop begins and stores them in the precomputation tensor. This approach eliminates redundant calculations at each stage of the NTT, thereby reducing computational overhead. Additionally, as in FNTT-DS, a vectorized approach is employed for the butterfly operations, processing the even and odd components in parallel through batched tensor operations.
 
\section{Experiment}

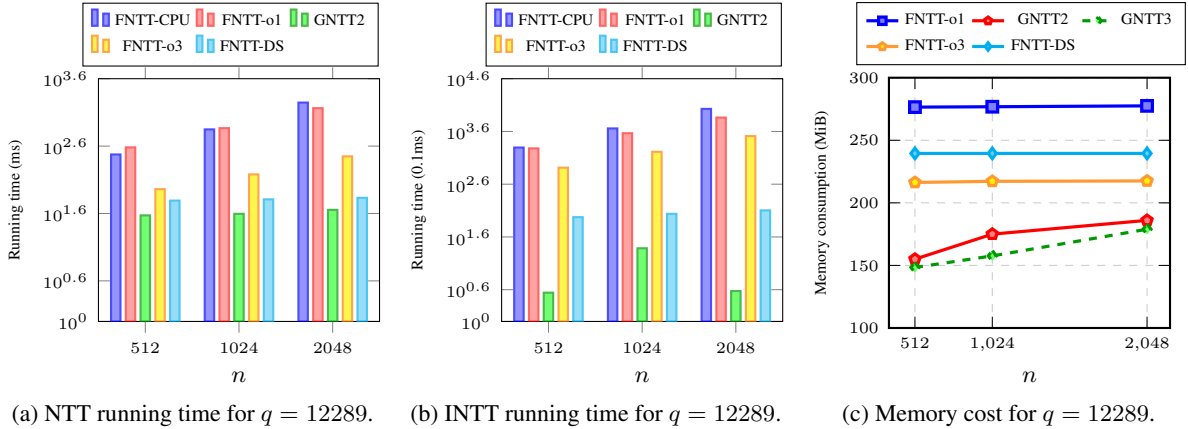
\begin{figure*}[t!]
	\centering
        \subfloat[{\small NTT running time for $q=12289$.}]{
		\begin{tikzpicture}
    \begin{axis}[
        ybar, 
        symbolic x coords={$512$, $1024$, $2048$},  
        xtick=data,  
        ymode=log,  
        xticklabel style={font=\tiny},
        yticklabel style={font=\tiny},
        ylabel={\tiny Running time (ms)},  
        xlabel={\small $n$},  
        width=0.33\linewidth,
        height=4.8cm,
        bar width=0.124cm,  
        enlarge x limits=0.25,  
        ymin=1,  
        ymax=4000,
        ytick={1, 4, 40, 400, 4000}, 
        ylabel near ticks,
        legend style={
            at={(0.47,1.05)}, 
            anchor=south,    
            legend columns=3, 
            draw=black,      
            fill=white,      
            font=\tiny,
            legend image post style={scale=0.7} 
        },
    ]

    \addplot[fill=blue!60,draw=blue,thick,opacity=0.7]
     coordinates {($512$, 300.368) ($1024$, 707.702) ($2048$, 1769.024)};  

    \addplot[fill=red!60,draw=red,thick,opacity=0.6]
     coordinates {($512$, 382.759) ($1024$, 737.988) ($2048$, 1461.154)};  

    \addplot[fill=green!80,draw=green!60!black,thick,opacity=0.7]
     coordinates {($512$, 37.493) ($1024$, 39.51) ($2048$, 45.258)};   

     \addplot[fill=yellow,draw=orange,thick,opacity=0.7]
    coordinates {($512$, 91.889) ($1024$, 152.051) ($2048$, 281.401)};  

    \addplot[fill=cyan!50,draw=cyan,thick,opacity=0.7]
    coordinates {($512$, 62.095) ($1024$, 64.79) ($2048$, 68.304)};

    \legend{FNTT-CPU, FNTT-o1, GNTT2, FNTT-o3, FNTT-DS}
    \end{axis}
\end{tikzpicture}
		\label{fig: ntt12289}	
	}	
        \subfloat[{\small INTT running time for $q=12289$.}]{
		\begin{tikzpicture}
    \begin{axis}[
        ybar, 
        symbolic x coords={$512$, $1024$, $2048$},  
        xtick=data,  
        ymode=log,  
        xticklabel style={font=\tiny},
        yticklabel style={font=\tiny},
        ylabel={\tiny Running time (0.1ms)},  
        xlabel={\small $n$},
        width=0.33\linewidth,
        height=4.8cm,
        bar width=0.124cm,  
        enlarge x limits=0.25,  
        ymin=1,  
        ymax=40000,
        ytick={1, 4, 40, 400, 4000, 40000}, 
        ylabel near ticks,
        legend style={
            at={(0.47,1.05)}, 
            anchor=south,    
            legend columns=3, 
            draw=black,      
            fill=white,      
            font=\tiny,
            legend image post style={scale=0.7} 
        },
    ]

    \addplot[fill=blue!60,draw=blue,thick,opacity=0.7]
     coordinates {($512$, 1980.72) ($1024$, 4561.33) ($2048$, 10702.09)};  

    \addplot[fill=red!60,draw=red,thick,opacity=0.6]
     coordinates {($512$, 1909.34) ($1024$, 3701.59) ($2048$, 7332.82)};  

    \addplot[fill=green!80,draw=green!60!black,thick,opacity=0.7]
     coordinates {($512$, 3.51) ($1024$, 24.38) ($2048$, 3.77)};   

     \addplot[fill=yellow,draw=orange,thick,opacity=0.7]
    coordinates {($512$, 822.89) ($1024$, 1644.04) ($2048$, 3266.27)};  

    \addplot[fill=cyan!50,draw=cyan,thick,opacity=0.7]
    coordinates {($512$, 94.86) ($1024$, 109.6) ($2048$, 127.52)};

    \legend{FNTT-CPU, FNTT-o1, GNTT2, FNTT-o3, FNTT-DS}
    \end{axis}
\end{tikzpicture}
		\label{fig: intt12289}	
	}
        \subfloat[{\small Memory cost for $q=12289$.}]{
		\begin{tikzpicture}
    \begin{axis}[
        width=0.33\linewidth,
        height=4.9cm,
        xlabel={\small $n$}, 
        ylabel={\tiny Memory consumption (MiB)}, 
        yticklabel style={font=\tiny},
        xticklabel style={font=\tiny},
        grid=major, 
        grid style={dashed, gray!40}, 
        legend style={
            at={(0.47,1.05)}, 
            anchor=south,    
            legend columns=3,
            font=\tiny,
            legend image post style={scale=0.7}}, 
        xmin=340, xmax=2200, 
        ymin=100, ymax=300, 
        xtick={512, 1024, 2048}, 
        ytick={100, 150, 200, 250, 300}, 
        tick align=inside, 
        tick style={black}, 
        axis line style={thick}, 
        every axis plot/.style={thick}, 
        ylabel near ticks,
        xlabel near ticks,
        ]
        
        \addplot[
            color=blue, 
            mark=square*, 
            mark options={fill=blue!50},
            line width=1.2pt
        ] coordinates {
            (512, 276.5) (1024, 276.8) (2048, 277.5)
        };
        \addlegendentry{FNTT-o1}

        \addplot[
            color=red, 
            mark=pentagon*, 
            mark options={fill=red!50},
            mark size=2pt,
            line width=1.2pt
        ] coordinates {
           (512, 155) (1024, 175) (2048, 186)
        };
        \addlegendentry{GNTT2}

        \addplot[
            dashed,
            color=green!60!black, 
            mark=diamond*, 
            mark options={fill=green!60},
            line width=1.2pt
        ] coordinates {
            (512, 148.4) (1024, 157.6) (2048, 178.9)
        };
        \addlegendentry{GNTT3}

        \addplot[
            color=orange!80, 
            mark=pentagon*, 
            mark options={fill=yellow},
            line width=1.2pt
        ] coordinates {
            (512, 216.3) (1024, 217.2) (2048, 217.5)
        };
        \addlegendentry{FNTT-o3}

        \addplot[
            color=cyan, 
            mark=diamond*, 
            mark options={fill=cyan!50},
            line width=1.2pt
        ] coordinates {
            (512, 239.5) (1024, 239.5) (2048, 239.5)
        };
        \addlegendentry{FNTT-DS}
    \end{axis}
\end{tikzpicture}
		\label{fig: mem12289}	
	}
        
\vspace{7pt}

    \caption{Evaluation results based on parameters $q$ and $n$ that satisfy PQC for all algorithms.}
    \label{fig:evaPQC}
\end{figure*}

\subsection{Implementation and Setup}

We conduct our experiments on a machine equipped with an NVIDIA A10 GPU, an 8-core CPU, and 30 GiB of RAM, running Ubuntu 22.04. The CPU-friendly Fast-NTT (FNTT-CPU) code is implemented using open-source code \footnote{https://github.com/Jyun-Neng/NTT}. We evaluate FNTT-CPU, GNTT4, FNTT-o1, FNTT-o3, and FNTT-DS from two perspectives: algorithm execution time and memory consumption. The baseline is composed of GNTT and FNTT-CPU. Additionally, it is important to note that due to potential issues in GPU-friendly code generated directly by LLMs, which may lead to computational or compilation errors, the code used for experimental benchmarking has been manually adjusted or prompted for correction by the LLMs. These adjustments do not involve the core algorithm principles.

\subsection{Metrics}
The algorithm execution time consists of two stages: NTT and INTT. Memory consumption is measured based on the entire function call process, excluding the precomputation of matrix generation.
The comprehensive speedup $S$ is computed as follows:
\begin{equation}
    S_{l} = \frac{ time_{\text{NTT}}^{c} + time_{\text{INTT}}^{c}}{time_{\text{NTT}}^{l} + time_{\text{INTT}}^{l}},
\end{equation}
where $c$ represents FNTT-CPU, and $l$ denotes the GPU-friendly algorithm.
The final score $s$ is calculated as follows:
\begin{equation}
    s = \frac{S_{l}}{S_{\text{baseline}}} + \frac{M_{\text{baseline}}}{M_{l}},
\end{equation}
where $M$ is the memory consumption.

Moreover, the parameters used in the experiments, especially $n$ and $q$, are selected based on the recommendations from both NIST Post-Quantum Cryptography (PQC) standards \footnote{https://csrc.nist.gov/projects/post-quantum-cryptography} and HE literature, although the parameter choices vary between these fields. For PQC, the parameter selection follows the standards outlined in \citep{NumberTheoreticTransformAndItsApplications}. On the other hand, for HE, the parameters are chosen based on research \citep{NTTGen, EfficientNTTOnGPU} focused on optimizing NTT for efficient implementation. The experimental data are drawn from multiple sets of results, emphasizing the data points that converge towards the median value.

\begin{table*}[t!]
\centering
\renewcommand{\arraystretch}{1.3}
\setlength{\tabcolsep}{4.5pt}
\scalebox{0.85}{  
  \begin{tabular}{lccccccc}
    \toprule
  \textbf{Algorithm}  & \textbf{Parameter $q$} & \textbf{Parameter $n$} & \textbf{NTT time (s)} & \textbf{INTT time (s)} & \textbf{Speedup} & \textbf{Memory (MiB)} & \textbf{Score}\\
    \midrule
 
 FNTT-CPU  &  $2^{23} - 2^{13} + 1$ & 2048 & 2.365035 & 1.433533 & 1.00 & 1.3 & - \\
 GNTT4  &  $2^{23} - 2^{13} + 1$ & 2048 & 0.045315 & 0.026255 & 53.07 & 222.7 & - \\
 \hdashline
 FNTT-o1  &  $2^{23} - 2^{13} + 1$ & 2048 & 1.516539 & 0.757572 &1.67 & 278.0 & 0.83\\   
 FNTT-o3  &  $2^{23} - 2^{13} + 1$ & 2048 & 0.287637 & 0.336224 & 6.09 & \colorbox{green!25}{\textbf{217.5}} & 1.14\\
 FNTT-DS  &  $2^{23} - 2^{13} + 1$ & 2048 & \colorbox{green!25}{\textbf{0.071202}} & \colorbox{red!25}{\textbf{0.014727}}  & \colorbox{green!25}{\textbf{44.21}} & 236.6 & 1.77 \\
\arrayrulecolor[gray]{0.5} \midrule
 FNTT-CPU  &  $2^{30} - 2^{18} + 1$ & 1024 & 0.924408 & 0.585499 & 1.00 & 0.5  &  -\\
 GNTT4  &  $2^{30} - 2^{18} + 1$ & 1024 & 0.040413 & 0.028420 & 21.94& 217.1 & -\\
 \arrayrulecolor{black} \hdashline
 FNTT-o1  &  $2^{30} - 2^{18} + 1$ & 1024 & 0.762433 & 0.379534  & 1.32 & 277.5 & 0.84\\
 FNTT-o3  &  $2^{30} - 2^{18} + 1$ & 1024 & 0.154871 &  0.166861 & 4.69 & \colorbox{green!25}{\textbf{217.2}} & 1.21\\
 FNTT-DS  &  $2^{30} - 2^{18} + 1$ & 1024 & \colorbox{green!25}{\textbf{0.067743}} & \colorbox{red!25}{\textbf{0.013848}}  & \colorbox{green!25}{\textbf{18.51}} & 236.5 & 1.76\\
 \arrayrulecolor[gray]{0.5} \midrule
 FNTT-CPU  &  $2^{30} - 2^{18} + 1$ & 2048 & 2.370426 & 1.432466 & 1.00 & 1.3 & -\\
 GNTT4  &  $2^{30} - 2^{18} + 1$ & 2048 & 0.045228 & 0.026296 &53.17 & 226.0 & -\\
 \arrayrulecolor{black} \hdashline
 FNTT-o1  &  $2^{30} - 2^{18} + 1$ & 2048 & 1.521742 & 0.761007 & 1.67 & 278.0 & 0.84\\
 FNTT-o3  &  $2^{30} - 2^{18} + 1$ & 2048 & 0.285375 & 0.33635 & 6.12 &  \colorbox{red!25}{\textbf{217.6}} & 1.15\\
 FNTT-DS  &  $2^{30} - 2^{18} + 1$ & 2048 & \colorbox{green!25}{\textbf{0.070627}} & \colorbox{red!25}{\textbf{0.015392}} & \colorbox{green!25}{\textbf{44.21}} & 236.6 & 1.79\\
    
\bottomrule
\end{tabular}
}
\caption{Evaluation results based on large parameters $q$ and $n$ that satisfy HE for all algorithms. The values where the algorithm achieves optimal performance relative to others generated by LLMs are highlighted in \colorbox{green!25}{\textbf{green}}, while values that outperform the baseline are highlighted in \raisebox{0pt}[0pt][0pt]{\colorbox{red!25}{\textbf{red}}.}}
\label{tab:Evaluation1}
\end{table*}

\subsection{Main Results}

We discuss the experimental results separately based on the variations in experimental parameters.
For experiments with PQC parameters, the algorithm execution time and memory consumption benchmarking results are shown in \autoref{fig:evaPQC}. Due to space limitations, additional experimental results can be found in \autoref{sec: AppendixEva}. Overall, FNTT-o1 exhibits the poorest performance due to its unoptimized execution speed (even worse than the original FNTT-CPU) and the highest memory consumption. Although FNTT-DS significantly outperforms FNTT-o3 in execution time, it still falls far short of the baseline. FNTT-o3 shows better memory cost than FNTT-DS, yet fails to surpass the baseline either. The comprehensive analysis reveals that GNTT2 maintains a substantial advantage over algorithms generated by LLMs, achieving a speedup of $s=62.22$ when $n=2048$ and $q=12289$. Furthermore, GNTT3 offers additional memory reduction. The precomputation test results, presented in \autoref{fig:Precomputation} in \autoref{sec: AppendixEva}, indicate that GNTT2 significantly reduces precomputation time compared to GNTT1. 

For experiments with HE parameters, the test results are summarized in \autoref{tab:Evaluation1}. Under large parameter configurations, GNTT4 shows noticeable degradation in speedup while demonstrating increased memory consumption. Notably, we observe the encouraging phenomenon that algorithms generated by LLMs can outperform the baseline. Among the comparative algorithms, FNTT-DS achieves a speedup approaching the baseline level, while FNTT-o3 maintains commendable performance in memory consumption.

Based on the above analysis, it can be concluded that the current LLMs each demonstrate distinct characteristics in their thinking, and can surpass our GNTT family in certain evaluation criteria. Specifically, DeepSeek-R1 clearly outperforms OpenAI o3-mini and o1. However, overall, they still exhibit considerable shortcomings.

\section{Related Work}
In recent years, GPU-friendly protocols proposed have been able to bring optimizations at different levels for PPML.
GForce \citep{Ng2021GForce} is an innovative framework to oblivious inference, integrating tailored cryptographic protocols with advancements in machine learning. It features a set of GPU-friendly protocols optimized for non-linear computations, leveraging GPU parallelism. However, GForce lacks generality for GPU acceleration. 
Considering the obstacles of realizing protocol-independent acceleration, Piranha \citep{Watson2022Piranha} provides a general-purpose scheme for MPC to utilize GPU acceleration.
Force \citep{Dai2023Force} is an efficient Four-Party PPML scheme on GPU, achieving significant performance improvements compared to other schemes such as Piranha by introducing novel protocols.
However, these schemes generally have defects in practicability.

\section{Conclusion}
To improve the efficiency of practical HE, we develop and optimize a GNTT family by exploiting PyTorch's fast matrix computation and precomputation. Our optimization ramps up the HE speed in record time. In addition, we explore the capability of LLMs to automatically generate GPU-friendly NTT code utilizing CPU-friendly Fast-NTT code. We carefully conduct a thorough algorithmic analysis of each model from multiple aspects, including GPU friendliness, precomputation, and the application of mathematical skills. Through extensive experiments, we demonstrate a number of interesting results, including one exhibiting DeepSeek-R1's superior GPU-friendly code generation capability over its competitors.

\section*{Limitations}
Due to the limitations of LLMs in GPU-friendly code generation tasks, our investigation is limited to NTT, the core mechanism of HE. Moreover, during the experiments, since the matrix computation function of PyTorch executed on GPU requires the input matrices or vectors to be in floating-point format. However, this requirement can introduce small computational errors, especially when the parameters $n$ and $q$ are quite large. Therefore, addressing the error introduced by floating-point computations is an important direction for future work.

\bibliography{latex/acl_latex}
\onecolumn
\newpage

\appendix

\section{Preliminaries}
\label{sec:notation}
\subsection{Terminology}
Many researchers do not distinguish the number theoretic transform itself from the FFT-based algorithms used to compute it. Following \citep{Satriawan2024BeginnerGuide}, we refer to the transform as the NTT (exhibiting quadratic complexity $O(n^2)$ when computed directly) and to any FFT-like procedure that performs this transform as the Fast-NTT (achieving quasi-linear complexity $O(n \log n)$).

\subsection{Notation}
The integer ring $\mathbb{Z}_q$ consists of the set of integers $\{0,1, \ldots, q-1\}$, equipped with addition and multiplication operations modulo $q$. Let the polynomial quotient ring $ \mathbb{Z}_q[x] / \phi(x)$ denote the set of equivalence classes of polynomials over $\mathbb{Z}_q$, where each polynomial is reduced modulo the irreducible polynomial $\phi(x)$. In the context of PQC and HE, the chosen rings are predominantly of the form $\mathbb{Z}_q[x]/(x^n + 1)$. Polynomial multiplication within the ring $\mathbb{Z}_q[x]/(x^n + 1)$ must be computed using negative wrapped convolution. Let $a(x) = \sum_{i=0}^{n-1} a_{i} x^i$ and $b(x) = \sum_{i=0}^{n-1} b_{i} x^i$ be polynomials of degree at most $n-1$ in the quotient ring $\mathbb{Z}_q[x] / (x^n + 1)$. The negative wrapped convolution $NWC(\cdot)$ is defined as:
\begin{equation}
\small
    NWC(a(x),b(x)) = \sum_{k=0}^{n-1} c_k x^k,
\end{equation}
where the coefficients $c_k$ are given by:
\begin{equation}
\small
    c_k = \left(\sum_{i=0}^k a_{i} \cdot b_{(k-i)}- \sum_{i=k+1}^{n-1} a_{i} \cdot b_{(k+n-i)}\right) \bmod q.
\end{equation}
This operation ensures that polynomial multiplication respects the structure of the quotient ring by incorporating reduction modulo $x^n + 1$.

Considering that optimization techniques will be employed in matrix computations, a polynomial $\boldsymbol{a}(x) = \sum_{i=0}^{n-1} a_i x^i$ in $\mathbb{Z}_q[x] / (x^n + 1)$ can be equivalently represented as a vector over $\mathbb{Z}_q$, denoted as $\boldsymbol{a} = [a_0, a_1, \ldots, a_{n-1}]$. Scalars are denoted by regular lowercase letters (e.g., $a$), polynomials by bold lowercase letters (e.g., $\boldsymbol{a}$), and vectors representing polynomials transformed via the Number Theoretic Transform are denoted with a prime symbol (e.g., $\boldsymbol{a^{\prime}}$ is the NTT representation of $\boldsymbol{a}$).

We define $\times$, $\odot$, and $\cdot$ to represent polynomial, element-wise vector, and integer multiplications, respectively. Throughout this paper, unless otherwise specified, $n$ denotes the degree of the polynomial ring, and $q$ denotes the prime modulus of its coefficients.

\section{Experiment Results}
\label{sec: AppendixEva}

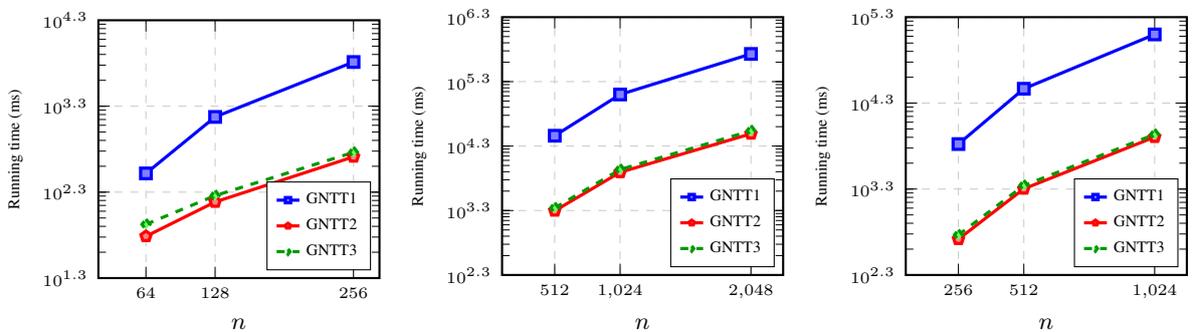
\begin{figure*}[!htbp]
	\centering
\subfloat[{\small Precomputation time for $q=7681$.}]{
		\begin{tikzpicture}
    \begin{axis}[
        width=0.33\linewidth,
        height=5cm,
        xlabel={\small $n$}, 
        ylabel={\tiny Running time (ms)}, 
        yticklabel style={font=\tiny},
        xticklabel style={font=\tiny},
        ymode = log,
        grid=major, 
        grid style={dashed, gray!40}, 
        legend pos=south east, 
        legend style={font=\tiny,
        legend image post style={scale=0.5}}, 
        xmin=20, xmax=280, 
        ymin=20, ymax=20000, 
        xtick={64, 128, 256}, 
        ytick={20, 200, 2000, 20000}, 
        tick align=inside, 
        tick style={black}, 
        axis line style={thick}, 
        every axis plot/.style={thick}, 
        ylabel near ticks,
        xlabel near ticks,
        ]
   
        \addplot[
            color=blue, 
            mark=square*, 
            mark options={fill=blue!50},
            line width=1.2pt
        ] coordinates {
            (64, 330.489) (128, 1499.978) (256, 6533.715)
        };
        \addlegendentry{GNTT1}

        \addplot[
            color=red, 
            mark=pentagon*, 
            mark options={fill=red!50},
            mark size=2pt,
            line width=1.2pt
        ] coordinates {
           (64, 61.23) (128, 153.682) (256, 515.246)
        };
        \addlegendentry{GNTT2}

        \addplot[
            dashed,
            color=green!60!black, 
            mark=diamond*, 
            mark options={fill=green!60},
            line width=1.2pt
        ] coordinates {
            (64, 84.298) (128, 182.167) (256, 580.603)
        };
        \addlegendentry{GNTT3}
    \end{axis}
\end{tikzpicture}
		\label{fig: matrix7681}	
	}
        \subfloat[{\small Precomputation time for $q=12289$.}]{
		\begin{tikzpicture}
    \begin{axis}[
        width=0.33\linewidth,
        height=5cm,
        xlabel={\small $n$}, 
        yticklabel style={font=\tiny},
        xticklabel style={font=\tiny},
        ylabel={\tiny Running time (ms)}, 
        ymode = log,
        grid=major, 
        grid style={dashed, gray!40}, 
        legend pos=south east, 
        legend style={font=\tiny, legend image post style={scale=0.5}}, % 图例样式
        xmin=100, xmax=2300, 
        ymin=200, ymax=2000000, 
        xtick={512, 1024, 2048}, 
        ytick={200, 2000, 20000, 200000, 2000000}, 
        tick align=inside, 
        tick style={black}, 
        axis line style={thick}, 
        every axis plot/.style={thick}, 
        ylabel near ticks,
        xlabel near ticks,
        ]
 
        \addplot[
            color=blue, 
            mark=square*, 
            mark options={fill=blue!50},
            line width=1.2pt
        ] coordinates {
            (512, 28931.892) (1024, 126295.654) (2048, 537723.467)
        };
        \addlegendentry{GNTT1}

        \addplot[
            color=red, 
            mark=pentagon*, 
            mark options={fill=red!50},
            mark size=2pt,
            line width=1.2pt
        ] coordinates {
           (512, 1979.617) (1024, 7839.182) (2048, 31005.553)
        };
        \addlegendentry{GNTT2}

        \addplot[
            dashed,
            color=green!60!black, 
            mark=diamond*, 
            mark options={fill=green!60},
            line width=1.2pt
        ] coordinates {
            (512, 2194.327) (1024, 8643.591) (2048, 34511.593)
        };
        \addlegendentry{GNTT3}
    \end{axis}
\end{tikzpicture}
		\label{fig: matrix12289}	
	}	
        \subfloat[{\small Precomputation time for $q=8380417$.}]{
		\begin{tikzpicture}
    \begin{axis}[
        width=0.33\linewidth,
        height=5cm,
        xlabel={\small $n$}, 
        yticklabel style={font=\tiny},
        xticklabel style={font=\tiny},
        ylabel={\tiny Running time (ms)}, 
        ymode = log,
        grid=major, 
        grid style={dashed, gray!40}, 
        legend pos=south east, 
        legend style={font=\tiny, legend image post style={scale=0.5}}, 
        xmin=50, xmax=1150, 
        ymin=200, ymax=200000, 
        xtick={256, 512, 1024}, 
        ytick={200, 2000, 20000, 200000}, 
        tick align=inside, 
        tick style={black}, 
        axis line style={thick}, 
        every axis plot/.style={thick}, 
        ylabel near ticks,
        xlabel near ticks,
        ]
        
        \addplot[
            color=blue, 
            mark=square*, 
            mark options={fill=blue!50},
            line width=1.2pt
        ] coordinates {
            (256, 6674.303) (512, 29372.257) (1024, 125933.016)
        };
        \addlegendentry{GNTT1}

        \addplot[
            color=red, 
            mark=pentagon*, 
            mark options={fill=red!50},
            mark size=2pt,
            line width=1.2pt
        ] coordinates {
           (256, 521.704) (512, 2012.047) (1024, 8044.069)
        };
        \addlegendentry{GNTT2}

        \addplot[
            dashed,
            color=green!60!black, 
            mark=diamond*, 
            mark options={fill=green!60},
            line width=1.2pt
        ] coordinates {
            (256, 584.681) (512, 2199.413) (1024, 8640.689)
        };
        \addlegendentry{GNTT3}
    \end{axis}
\end{tikzpicture}
		\label{fig: matrix8380417}	
	}
    \caption{Evaluation results of the precomputation time for GNTT algorithms.}
	\label{fig:Precomputation}
\end{figure*}

\begin{figure*}
	\centering
        \subfloat[{\small NTT running time for $q=7681$.}]{
		\begin{tikzpicture}
    \begin{axis}[
        ybar, 
        symbolic x coords={$n=64$, $n=128$, $n=256$},  ymode=log, 
        xtick=data,  
        xticklabel style={font=\small},
        yticklabel style={font=\small},
        ylabel={\small Running time (ms)},    
        width=0.45\linewidth,
        height=6.2cm,
        bar width=0.18cm,  
        enlarge x limits=0.25,  
        ymin=1,  
        ymax=400,
        ytick={1, 4, 40, 400}, 
        ylabel near ticks,
        legend style={
            at={(0.5,1.05)}, 
            anchor=south,    
            legend columns=3, 
            draw=black,      
            fill=white,    
            font=\scriptsize,
            legend image post style={scale=0.8} 
        },
    ]

    \addplot[fill=blue!60,draw=blue,thick,opacity=0.7]
     coordinates {($n=64$, 20.19) ($n=128$, 50.199) ($n=256$, 121.189)};  

    \addplot[fill=red!60,draw=red,thick,opacity=0.6]
     coordinates {($n=64$, 69.46) ($n=128$, 113.481) ($n=256$, 206.315)};  

    \addplot[fill=green!80,draw=green!60!black,thick,opacity=0.7]
     coordinates {($n=64$, 38.882) ($n=128$, 41.145) ($n=256$, 39.375)};   

     \addplot[fill=yellow,draw=orange,thick,opacity=0.7]
    coordinates {($n=64$, 39.54) ($n=128$, 46.226) ($n=256$, 61.044)};

    \addplot[fill=cyan!50,draw=cyan,thick,opacity=0.7]
    coordinates {($n=64$, 55.137) ($n=128$, 56.628) ($n=256$, 58.39)};

    \legend{FNTT-CPU, FNTT-o1, GNTT2, FNTT-o3, FNTT-DS}
    \end{axis}
\end{tikzpicture}
		\label{fig: ntt7681}	
	}
        \hfill
        \subfloat[{\small NTT running time for $q=8380417$.}]{
		\begin{tikzpicture}
    \begin{axis}[
        ybar, 
        symbolic x coords={$n=256$, $n=512$, $n=1024$},  
        xtick=data,  
        ymode=log,  
        xticklabel style={font=\small},
        yticklabel style={font=\small},
        ylabel={\small Running time (ms)},  
        width=0.45\linewidth,
        height=6.2cm,
        bar width=0.18cm,  
        enlarge x limits=0.25,  
        ymin=1,  
        ymax=4000,
        ytick={1, 4, 40, 400, 4000}, 
        ylabel near ticks,
        legend style={
            at={(0.5,1.05)}, 
            anchor=south,    
            legend columns=3, 
            draw=black,      
            fill=white,      
            font=\scriptsize, 
            legend image post style={scale=0.8} 
        },
    ]

    \addplot[fill=blue!60,draw=blue,thick,opacity=0.7]
     coordinates {($n=256$, 125.413) ($n=512$, 299.832) ($n=1024$, 727.731)};  

    \addplot[fill=red!60,draw=red,thick,opacity=0.6]
     coordinates {($n=256$, 204.177) ($n=512$, 381.066) ($n=1024$, 762.482)};  

    \addplot[fill=green!80,draw=green!60!black,thick,opacity=0.7]
     coordinates {($n=256$, 39.533) ($n=512$, 38.055) ($n=1024$, 40.129)};   

     \addplot[fill=yellow,draw=orange,thick,opacity=0.7]
    coordinates {($n=256$, 60.873) ($n=512$, 90.361) ($n=1024$, 152.783)};  
    
    \addplot[fill=cyan!50,draw=cyan,thick,opacity=0.7]
    coordinates {($n=256$, 60.169) ($n=512$, 61.552) ($n=1024$, 64.263)};

    \legend{FNTT-CPU, FNTT-o1, GNTT2, FNTT-o3, FNTT-DS}

    \end{axis}
\end{tikzpicture}
		\label{fig: ntt8380417}	
	}
\caption{Comparison of the evaluation results for NTT running time.}
	\label{fig:ntt}
\end{figure*}
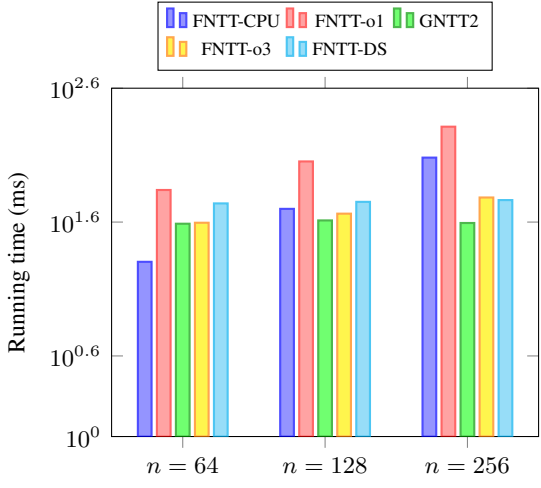
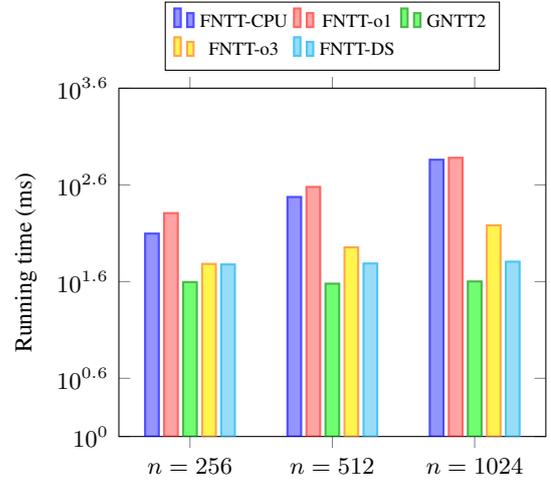

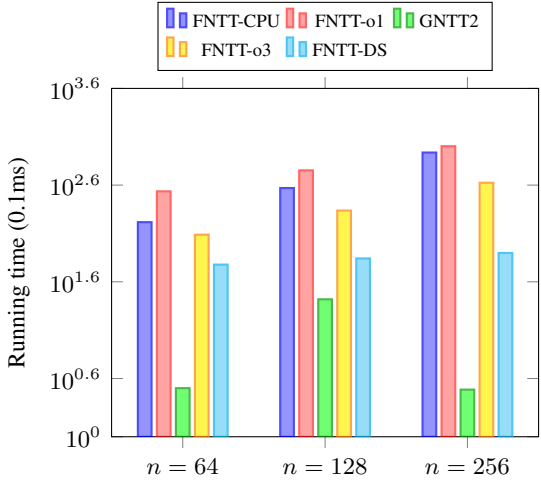
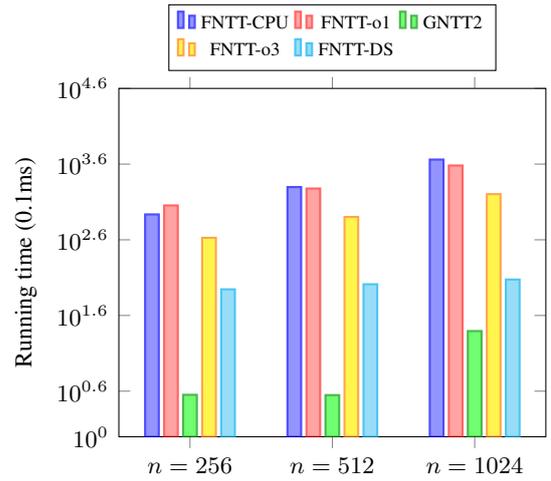
\begin{figure*}
        \subfloat[{\small INTT running time for $q=7681$.}]{
		\begin{tikzpicture}
    \begin{axis}[
        ybar, 
        symbolic x coords={$n=64$, $n=128$, $n=256$}, 
        xtick=data,  
        xticklabel style={font=\small},
        yticklabel style={font=\small},
        ylabel={\small Running time (0.1ms)},  
        ymode=log, 
        width=0.45\linewidth,
        height=6.2cm,
        bar width=0.18cm,  
        enlarge x limits=0.25,  
        ymin=1,  
        ymax=4000,
        ytick={1, 4, 40, 400, 4000}, 
        ylabel near ticks,
        legend style={
            at={(0.5,1.05)}, 
            anchor=south,    
            legend columns=3, 
            draw=black,      
            fill=white,      
            font=\scriptsize,
            legend image post style={scale=0.8} 
        },
    ]

    \addplot[fill=blue!60,draw=blue,thick,opacity=0.7]
     coordinates {($n=64$, 165.63) ($n=128$, 373.38) ($n=256$, 868.84)};  

    \addplot[fill=red!60,draw=red,thick,opacity=0.6]
     coordinates {($n=64$, 344.77) ($n=128$, 568.38) ($n=256$, 1009.35)};  

    \addplot[fill=green!80,draw=green!60!black,thick,opacity=0.7]
     coordinates {($n=64$, 3.18) ($n=128$, 26.38) ($n=256$, 3.07)};   

     \addplot[fill=yellow,draw=orange,thick,opacity=0.7]
    coordinates {($n=64$, 122.9) ($n=128$, 218.27) ($n=256$, 422.96)};

    \addplot[fill=cyan!50,draw=cyan,thick,opacity=0.7]
    coordinates {($n=64$, 60.27) ($n=128$, 69.78) ($n=256$, 79.54)};

    \legend{FNTT-CPU, FNTT-o1, GNTT2, FNTT-o3, FNTT-DS}
    \end{axis}
\end{tikzpicture}
		\label{fig: intt7681}	
	}
        \hfill
        \subfloat[{\small INTT running time for $q=8380417$.}]{
		\begin{tikzpicture}
    \begin{axis}[
        ybar, 
        symbolic x coords={$n=256$, $n=512$, $n=1024$},  
        xtick=data,  
        ymode=log,  
        xticklabel style={font=\small},
        yticklabel style={font=\small},
        ylabel={\small Running time (0.1ms)},   
        width=0.45\linewidth,
        height=6.2cm,
        bar width=0.18cm,  
        enlarge x limits=0.25,  
        ymin=1,  
        ymax=40000,
        ytick={1, 4, 40, 400, 4000, 40000}, 
        ylabel near ticks,
        legend style={
            at={(0.5,1.05)}, 
            anchor=south,    
            legend columns=3, 
            draw=black,      
            fill=white,      
            font=\scriptsize,
            legend image post style={scale=0.7} 
        },
    ]

    \addplot[fill=blue!60,draw=blue,thick,opacity=0.7]
     coordinates {($n=256$, 865.96) ($n=512$, 1989.84) ($n=1024$, 4593.17)};  

    \addplot[fill=red!60,draw=red,thick,opacity=0.6]
     coordinates {($n=256$, 1137.27) ($n=512$, 1904.05) ($n=1024$, 3833.99)};  

    \addplot[fill=green!80,draw=green!60!black,thick,opacity=0.7]
     coordinates {($n=256$, 3.58) ($n=512$, 3.55) ($n=1024$, 24.9)};   

     \addplot[fill=yellow,draw=orange,thick,opacity=0.7]
    coordinates {($n=256$, 425.05) ($n=512$, 801.24) ($n=1024$, 1604.64)};  

    \addplot[fill=cyan!50,draw=cyan,thick,opacity=0.7]
    coordinates {($n=256$, 88.65) ($n=512$, 103.46) ($n=1024$, 119.45)};

    \legend{FNTT-CPU, FNTT-o1, GNTT2, FNTT-o3, FNTT-DS}
    \end{axis}
\end{tikzpicture}
		\label{fig: intt8380417}	
	}

    \caption{Comparison of the evaluation results for INTT running time.}
	\label{fig:ntt}
\end{figure*}

\begin{figure*}
	\centering
        \subfloat[{\small Memory cost for $q=7681$.}]{
		\begin{tikzpicture}
    \begin{axis}[
        width=0.45\linewidth,
        height=6.2cm,
        xlabel={\small $n$}, 
        xticklabel style={font=\small},
        ylabel={\small Memory consumption (MiB)}, 
        yticklabel style={font=\small},
        grid=major, 
        grid style={dashed, gray!40}, 
        legend style={
            at={(0.5,1.05)}, 
            anchor=south,    
            legend columns=3,
            font=\scriptsize,
            legend image post style={scale=0.8}}, 
        xmin=30, xmax=280, 
        ymin=100, ymax=300, 
        xtick={64, 128, 256}, 
        ytick={100, 150, 200, 250, 300}, 
        tick align=inside, 
        tick style={black}, 
        axis line style={thick}, 
        every axis plot/.style={thick}, 
        ylabel near ticks,
        xlabel near ticks,
        ]
       
        \addplot[
            color=blue, 
            mark=square*, 
            mark options={fill=blue!50},
            line width=1.2pt
        ] coordinates {
            (64, 276.1) (128, 276) (256, 276.2)
        };
        \addlegendentry{FNTT-o1}

        \addplot[
            color=red, 
            mark=pentagon*, 
            mark options={fill=red!50},
            mark size=2pt,
            line width=1.2pt
        ] coordinates {
           (64, 163.2) (128, 172.3) (256, 163.2)
        };
        \addlegendentry{GNTT2}

        \addplot[
            dashed,
            color=green!60!black, 
            mark=diamond*, 
            mark options={fill=green!60},
            line width=1.2pt
        ] coordinates {
            (64, 165.3) (128, 174.5) (256, 164.7)
        };
        \addlegendentry{GNTT3}

        \addplot[
            color=orange!80, 
            mark=pentagon*, 
            mark options={fill=yellow},
            line width=1.2pt
        ] coordinates {
            (64, 216.4) (128, 216.4) (256, 216.3)
        };
        \addlegendentry{FNTT-o3}

        \addplot[
            color=cyan, 
            mark=diamond*, 
            mark options={fill=cyan!50},
            line width=1.2pt
        ] coordinates {
            (64, 239.4) (128, 239.9) (256, 239.7)
        };
        \addlegendentry{FNTT-DS}
        
    \end{axis}
\end{tikzpicture}
		\label{fig: mem7681}	
	}
        \hfill
        \subfloat[{\small Memory cost for $q=8380417$.}]{
		\begin{tikzpicture}
    \begin{axis}[
        width=0.45\linewidth,
        height=6.2cm,
        xlabel={\small $n$}, 
        ylabel={\small Memory consumption (MiB)}, 
        xticklabel style={font=\small},
        yticklabel style={font=\small},
        grid=major, 
        grid style={dashed, gray!40}, 
        legend style={
            at={(0.5,1.05)}, 
            anchor=south,    
            legend columns=3,
            font=\scriptsize,
            legend image post style={scale=0.8}}, 
        xmin=180, xmax=1110, 
        ymin=100, ymax=300, 
        xtick={256, 512, 1024}, 
        ytick={100, 150, 200, 250, 300}, 
        tick align=inside, 
        tick style={black}, 
        axis line style={thick}, 
        every axis plot/.style={thick}, 
        ylabel near ticks,
        xlabel near ticks,
        ]
        
        \addplot[
            color=blue, 
            mark=square*, 
            mark options={fill=blue!50},
            line width=1.2pt
        ] coordinates {
            (256, 276.3) (512, 276.4) (1024, 276.7)
        };
        \addlegendentry{FNTT-o1}

        \addplot[
            color=red, 
            mark=pentagon*, 
            mark options={fill=red!50},
            mark size=2pt,
            line width=1.2pt
        ] coordinates {
           (256, 163.3) (512, 155.2) (1024, 174.8)
        };
        \addlegendentry{GNTT2}

        \addplot[
            dashed,
            color=green!60!black, 
            mark=diamond*, 
            mark options={fill=green!60},
            line width=1.2pt
        ] coordinates {
            (256, 164.8) (512, 148.4) (1024, 165.6)
        };
        \addlegendentry{GNTT3}

        \addplot[
            color=orange!80, 
            mark=pentagon*, 
            mark options={fill=yellow},
            line width=1.2pt
        ] coordinates {
            (256, 216.3) (512, 216) (1024, 216.4)
        };
        \addlegendentry{FNTT-o3}

        \addplot[
            color=cyan, 
            mark=diamond*, 
            mark options={fill=cyan!50},
            line width=1.2pt
        ] coordinates {
            (256, 246.4) (512, 253.7) (1024, 253.2)
        };
        \addlegendentry{FNTT-DS}
        
    \end{axis}
\end{tikzpicture}
		\label{fig: mem8380417}	
	}
    
    \caption{Comparison of the evaluation results for memory consumption.}
	\label{fig:memory}
\end{figure*}
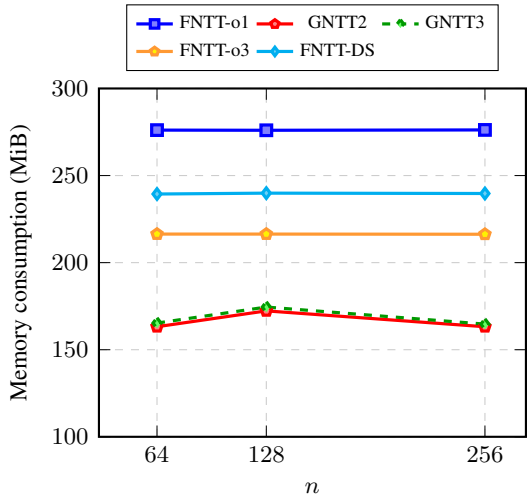
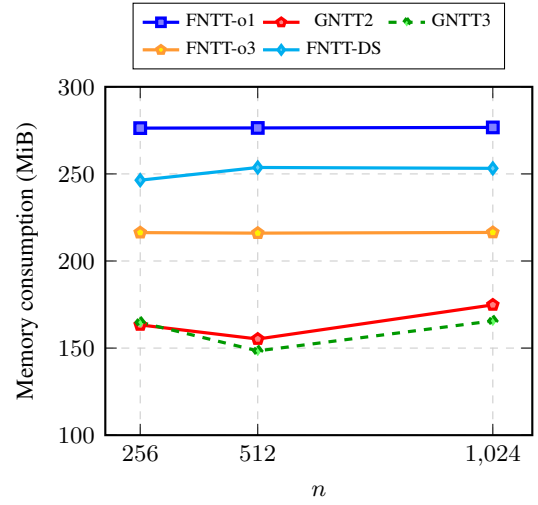

\end{document}